# Exact Solution of Chandrasekhar's H Function For Isotropic Case

Fikret Anlı

Email: anli@ksu.edu.tr, ORICID: 0000-0001-8357-340X

Kahramanmaraş Sütçü İmam University, Faculty of Sciences
Department of Physics Kahramanmaraş/Turkey

**Abstract:** In this work, we obtained the exact solution of Chandrasekhar's H function with isotropic scattering. The H function is a nonlinear integral equation that has long kept scientists busy in the field of radiative transfer theory. In this study, we first obtained the differential form of the integral equation because the solution would be easy. We used the classical integral technique to construct the differential form. By solving the differential equation we obtained exact solution of the H function. We compared the numerical results of exact solution with the numerical results obtained by Chandrasekhar in his book. We introduced them in the table to see the differences.

**Keywords:** Radiative Transfer; Integral Equations; Chandrasekhar H Function; Moments of H Function; Light Scattering in the Atmosphere of Planets

## 1. Introduction

As will be explained in detail in the next section, the H Function is an important function for radiative transfer in the atmosphere of planets (Chandrasekhar 1960) and its solution is a necessity. The fact that the equation is a nonlinear integral equation makes the solution very difficult. So far, no exact analytical solution of the H function has been found. In cases where an exact analytical solution cannot be found, numerical solutions must be sought. What kind of approach will be used and what kind of method will be applied in solving this type of nonlinear equations is also an important issue.

Approximate solutions of the H function are realized by proposing an approximate analytical function and defining an algorithm that performs fast solution of the integral equation representing the H function (Jablonski 2015). Seeking an analytical solution was carried out using the mean value theorem for the integral in the integral equation and reasonable results have been obtained (Vukanic 2007). The H function and its moments are calculated with the help of a computer algorithm that gives compact, fast, reliable and high-digit numerical results (Kawabata 2022). A new approach using shifted Legendre polynomials is applied for the numerical solution of the nonlinear integral equation for the Chandrasekhar H-function for isotropic scattering and the results are in good agreement with the literature and the first derivative of the H-function is also given numerically (Anlı and Öztürk 2021). By using an exponential approach to the integral equation, a new integral equation similar to the H function was obtained and both the solution of the H function and its moments were calculated numerically (Öztürk and Anli 2023). The H-function and its derivative have been tabulated to describe light reflected from the surface of celestial bodies and to understand the radiation transport in spheres (Hapke 1981; Hulst 1980). In conclusion, thousands of articles have been written on this subject, but none have been found to have a complete solution.

In the second part, we obtained the differential equation of the H function by a very comprehensive analysis on the integral equation of the H function. In the third part, we presented the solution of the differential equation by two different methods to verify the accuracy of the processes performed. In the last section, we presented the results we obtained and made some explanations.

## 2. Evolution and Analysis on the H function

The nonlinear integral equation for the Chandrasekhar's H function with isotropic scattering and for non-conservative case is given by

$$H(\mu,\omega)=1+\frac{\omega\mu}{2}H(\mu,\omega)\int_0^1\frac{H(\mu',\omega)}{\mu+\mu'}d\mu',\quad 0\le(\omega,\mu)\le 1 \tag{1}$$

where $\mu$ is the directional variable and $\omega$ is the single-scattering albedo As shown in Eq.(1), the H function is equal to unity for both case $\mu=0$ and $\omega=0$. We write the above equation in a more convenient form for use in proceeding processes;

$$Z_0(\mu,\omega)=\frac{2}{\omega\mu}\left(1-\frac{1}{H(\mu,\omega)}\right)=\int_0^1\frac{H(\mu',\omega)}{\mu+\mu'}d\mu',\quad 0\le(\omega,\mu)\le 1 \tag{2}$$

Since finding the integral on the right side of Eq.(2) means solving the equation, our main goal here will be to do so. Therefore, we aim to simplify our procedures by rewriting Eq.(1) with the different variables;

$$H(\eta,\omega)=1+\frac{\omega\eta}{2}H(\eta,\omega)\int_0^1\frac{H(\mu',\omega)}{\eta+\mu'}d\mu',\quad 0\le(\omega,\eta)\le 1 \tag{3}$$

$$H(\mu',\omega)=1+\frac{\omega\mu'}{2}H(\mu',\omega)\int_0^1\frac{H(\eta,\omega)}{\mu'+\eta}d\eta,\quad 0\le(\omega,\mu')\le 1 \tag{4}$$

The reader might think that the variable increase or enrichment process we've applied here will create difficulties in solving the equation. However, on the contrary, it will be seen that it provides extraordinary ease of solution.

Let's multiply Eq.(3) by $d\eta/(\mu+\eta)$ and integrate over $\eta$ in the range of [0-1];

$$\int_0^1\frac{H(\eta,\omega)}{\mu+\eta}d\eta-\ln\left(1+\frac{1}{\mu}\right)=\frac{\omega}{2}\int_0^1\int_0^1\frac{\eta H(\eta,\omega)}{\mu+\eta}\frac{H(\mu',\omega)}{\eta+\mu'}d\eta d\mu',\quad 0\le(\omega,\mu)\le 1 \tag{5}$$

By the same way, let's multiply Eq.(4) by $d\mu'/(\mu+\mu')$ and integrate over $\mu'$ in the range of [0-1];

$$\int_0^1 \frac{H(\mu',\omega)}{\mu+\mu'}d\mu' - \ln\left(1+\frac{1}{\mu}\right) = \frac{\omega}{2}\int_0^1\int_0^1 \frac{\mu' H(\eta,\omega)}{\mu+\mu'}\frac{H(\mu',\omega)}{\eta+\mu'}d\eta d\mu', \quad 0 \le (\omega,\mu) \le 1 \tag{6}$$

When Eq.(5) and Eq.(6) are examined carefully, both equations are equal to each other, so we can write the following equation representing both equations;

$$Z_0(\mu,\omega) = \ln\left(1+\frac{1}{\mu}\right) + \frac{\omega}{2}Z_1(\mu,\omega), \qquad 0 \le (\omega,\mu) \le 1 \tag{7}$$

$Z_0(\mu,\omega)$ is defined in Eq.(2) and $Z_1(\mu,\omega)$ is defined as

$$Z_1(\mu,\omega) = \int_0^1\int_0^1 \frac{\mu' H(\mu',\omega)H(\eta,\omega)}{(\mu+\mu')(\eta+\mu')}d\mu' d\eta = \int_0^1\int_0^1 \frac{\eta H(\mu',\omega)H(\eta,\omega)}{(\mu+\eta)(\eta+\mu')}d\mu' d\eta, \tag{8}$$

Since the two integrals given in Eq.(8) are equal to each other, the difference of these two integrals must be equal to 0;

$$\int_0^1\int_0^1 \frac{(\mu'-\eta)H(\mu',\omega)H(\eta,\omega)}{(\mu+\mu')(\mu+\eta)(\eta+\mu')}d\mu' d\eta = 0, \qquad 0 \le (\omega,\mu) \le 1 \tag{9}$$

Using the $\eta/(\eta+\mu') = 1-\mu'/(\eta+\mu')$ identity in Eq.(9), we can derive the following result

$$\int_0^1\int_0^1 \frac{\mu' H(\mu',\omega)H(\eta,\omega)}{(\mu+\mu')(\mu+\eta)(\eta+\mu')}d\mu' d\eta = \frac{1}{2}Z_0^2(\mu,\omega), \qquad 0 \le (\omega,\mu) \le 1 \tag{10}$$

From Eq.(9) and Eq.(10) one can easily derive the following equation;

$$\int_0^1\int_0^1 \frac{\eta H(\mu',\omega)H(\eta,\omega)}{(\mu+\mu')(\mu+\eta)(\eta+\mu')}d\mu' d\eta = \frac{1}{2}Z_0^2(\mu,\omega), \qquad 0 \le (\omega,\mu) \le 1 \tag{11}$$

Eq.(10 ) and Eq.(11) are very useful equations those we can use to achieve our goal. When we use the identity $\mu/(\eta+\mu) = 1-\eta/(\eta+\mu)$ after multiplying Eq.(10) by $\mu$, we can obtain the following equation;

$$Z_1(\mu,\omega) = \frac{\mu}{2}Z_0^2(\mu,\omega) + Z_2(\mu,\omega), \qquad 0 \le (\omega,\mu) \le 1 \tag{12}$$

where $Z_2(\mu,\omega)$ function is defined as

$$Z_2(\mu,\omega) = \int_0^1\int_0^1 \frac{\eta\mu' H(\mu',\omega)H(\eta,\omega)}{(\mu+\mu')(\mu+\eta)(\eta+\mu')}d\mu' d\eta, \quad 0 \le (\omega,\mu) \le 1 \tag{13}$$

when we substitute the first integral in Eq.(8) and Eq.(13) into Eq.(12), take the derivative of each term with respect to $\mu$, and rearrange the result, we find the following equation;

$$\int_0^1\int_0^1 \frac{\mu\mu' H(\mu',\omega)H(\eta,\omega)d\mu' d\eta}{(\mu+\mu')^2(\mu+\eta)(\eta+\mu')} = -\frac{1}{2}\frac{\partial}{\partial\mu}\left(\mu Z_0^2(\mu,\omega)\right) + \int_0^1\int_0^1 \frac{\eta\mu' H(\mu',\omega)H(\eta,\omega)d\mu' d\eta}{(\mu+\mu')(\mu+\eta)^2(\eta+\mu')}, \tag{14}$$

By the same way, when we substitute the second integral in Eq.(8) and Eq.(13) into Eq.(12), take the derivative of each term with respect to $\mu$ , and rearrange the result, we find the following equation;

$$\int_0^1\int_0^1 \frac{\mu\eta H(\mu',\omega)H(\eta,\omega)d\mu' d\eta}{(\mu+\mu')(\mu+\eta)^2(\eta+\mu')} = -\frac{1}{2}\frac{\partial}{\partial\mu}\left(\mu Z_0^2(\mu,\omega)\right) + \int_0^1\int_0^1 \frac{\eta\mu' H(\mu',\omega)H(\eta,\omega)d\mu' d\eta}{(\mu+\mu')^2(\mu+\eta)(\eta+\mu')}, \tag{15}$$

When we add Eq.(14) to Eq.(15) we get the following equation;

$$\mu\int_0^1\int_0^1 \left[\frac{\mu'}{(\mu+\mu')^2(\mu+\eta)} + \frac{\eta}{(\mu+\mu')(\mu+\eta)^2}\right]\frac{H(\mu',\omega)H(\eta,\omega)d\mu' d\eta}{(\eta+\mu')} =$$

$$-\frac{\partial}{\partial\mu}\left(\mu Z_0^2(\mu,\omega)\right) - \frac{\partial Z_2(\mu,\omega)}{\partial\mu}, \qquad 0 \le (\omega,\mu) \le 1 \tag{16}$$

When we evaluate the integrals in Eq.(16), we will have almost completed our operations to obtain the exact solution of H function. To do this, let's take the derivatives of Eq.(10) and Eq.(11) with respect to $\mu$ and add them together;

$$\int_0^1\int_0^1 \frac{\mu' H(\mu',\omega)H(\eta,\omega)}{(\mu+\mu')^2(\mu+\eta)(\eta+\mu')}d\mu' d\eta + \int_0^1\int_0^1 \frac{\mu' H(\mu',\omega)H(\eta,\omega)}{(\mu+\mu')(\mu+\eta)^2(\eta+\mu')}d\mu' d\eta +$$

$$\int_0^1\int_0^1 \frac{\eta H(\mu',\omega)H(\eta,\omega)}{(\mu+\mu')^2(\mu+\eta)(\eta+\mu')}d\mu' d\eta + \int_0^1\int_0^1 \frac{\eta H(\mu',\omega)H(\eta,\omega)}{(\mu+\mu')(\mu+\eta)^2(\eta+\mu')}d\mu' d\eta = -\frac{\partial}{\partial\mu}Z_0^2(\mu,\omega) \tag{17}$$

When Eq.(17) is examined carefully, it is seen that if the $\eta \leftrightarrow \mu'$ variable interchange is made in the fourth integral, it will be equal to the first integral then we decide that the first integral is equal to the fourth integral. Similarly, if the $\mu' \leftrightarrow \eta$ variable interchange is made in the second integral, it will be equal to third integral then we decide second integral is equal to third integral. Then, Eq.(17) can be rewritten as

$$\int_0^1\int_0^1 \frac{\mu' H(\mu',\omega)H(\eta,\omega)d\mu' d\eta}{(\mu+\mu')^2(\mu+\eta)(\eta+\mu')} + \int_0^1\int_0^1 \frac{\eta H(\mu',\omega)H(\eta,\omega)d\mu' d\eta}{(\mu+\mu')(\mu+\eta)^2(\eta+\mu')} = -\frac{1}{2}\frac{\partial}{\partial\mu}Z_0^2(\mu,\omega) \tag{18}$$

If we use Eq.(18) in Eq.(16), we obtain

$$\frac{\partial Z_2(\mu,\omega)}{\partial\mu} = -Z_0^2(\mu,\omega) - \frac{\mu}{2}\frac{\partial}{\partial\mu}Z_0^2(\mu,\omega), \qquad 0 \le (\omega,\mu) \le 1, \tag{19}$$

Let's differentiate Eq.(12) with respect to $\mu$ once and write Eq.(19) there;

$$\frac{\partial Z_1(\mu,\omega)}{\partial \mu} = -\frac{1}{2} Z_0^2(\mu,\omega), \qquad 0 \le (\omega,\mu) \le 1, \tag{20}$$

And finally, let's differentiate Eq.(7) with respect to $\mu$ once and write Eq.(20) there;

$$\frac{\partial Z_0(\mu,\omega)}{\partial \mu} + \frac{\omega}{4} Z_0^2(\mu,\omega) = \frac{d}{d\mu}\left( \ln\left(1+\frac{1}{\mu}\right)\right), \qquad 0 \le (\omega,\mu) \le 1, \tag{21}$$

$Z_0(\mu,\omega)$ is defined in Eq.(2), when we use the definition in Eq.(2) in Eq.(21), we obtain the following equation for the H function;

$$\frac{\partial H(\mu,\omega)}{\partial \mu} - \left( \frac{1}{2\mu} - \frac{\omega}{2(1+\mu)} \right) H^2(\mu,\omega) = -\frac{1}{2\mu}, \qquad 0 \le (\omega,\mu) \le 1, \tag{22}$$

Eq.(21) or Eq.(22) is the differential form of the Chandrasekhar H function in the integral equation given in Eq.(1). Eq.(21) and Eq.(22) are new findings, and although thousands of articles have been published on the subject, no one has yet found these differential forms. We derived Eq.(22) here for information purposes, we will not use it in our future operations, because Eq.(21) is more useful than it.

## 3. Moments and Solution of H function

The moments of the H function are defined as follows;

$$\alpha_n(\omega) = \int_0^1 \mu^n H(\mu,\omega)\, d\mu = \int_0^1 \mu'^n H(\mu',\omega)\, d\mu' = \int_0^1 \eta^n H(\eta,\omega)\, d\eta, \quad 0 \le \omega \le 1, \tag{23}$$

In the studies on this subject so far, an analytical functional expression could only be given for the moment $\alpha_0(\omega)$, but no analytical functional expression could be given for other $\alpha_n(\omega),\ n \ge 1$ moments. In this study, we will give all the $\alpha_n(\omega)$ moments explicitly functional form. For this purpose we can construct the following series from of Eq.(2)

$$Z_0(\mu,\omega) = \sum_{n=0}^{\infty} \frac{(-1)^n \alpha_n(\omega)}{\mu^{n+1}}, \qquad 0 \le (\omega,\mu) \le 1, \tag{24}$$

For this operation, we also need the formal series of $\ln(1+1/\mu)$ which is given by

$$\ln\left(1+\frac{1}{\mu}\right) = \sum_{n=0}^{\infty} \frac{(-1)^n}{(n+1)\mu^{n+1}}, \qquad 0 \le \mu \le 1 \tag{25}$$

Let's use Eq.(25) and Eq.(24) in Eq.(21) and expand it to series according to $1/\mu$ , analytical expressions for all moments can be found easily. We give the first few moments here;

$$\alpha_0(\omega) = \frac{2}{\omega}\left(1 - \sqrt{1-\omega}\right), \qquad 0 \le \omega \le 1 \tag{26}$$

$$\alpha_1(\omega) = \frac{1}{1+\sqrt{1-\omega}}, \qquad 0 \le \omega \le 1 \tag{27}$$

$$\alpha_2(\omega) = \frac{3\omega - 8(1-\sqrt{1-\omega})}{4(4-6\omega+(\omega-6)\sqrt{1-\omega})}, \qquad 0 \le \omega \le 1 \tag{28}$$

and recurrence relation for the moments can be given exactly as

$$4(n+1)\alpha_n(\omega) = 4 + \omega \sum_{m=0}^{n} \alpha_m(\omega)\alpha_{n-m}(\omega), \quad n \ge 0, \qquad 0 \le \omega \le 1, \tag{29}$$

functional forms of all the moments can be obtained using Eq.(29).

**Basic or Approximate Solution of H Function:** Eq.(21) and Eq.(22) are the first-order nonlinear differential equations, known as Riccati differential equation in mathematics. Both equations are not seen to be solvable easily. We prefer to use Eq.(21) as the equation whose solution we seek. Eq.(21) motivates us that there will be logarithmic functions in the differential equation. So we can start by expanding $Z_0(\mu,\omega)$ into a series of logarithmic functions (see Anli, Öztürk 2021);

$$\frac{1}{\mu+\mu'} = 2\sum_{n=0}^{\infty}(2n+1)(-1)^n Q_n(2\mu+1)P_n(2\mu'-1), \qquad 0 \le (\mu,\mu') \le 1 \tag{30}$$

where $P_n(2\mu'-1)$ and $Q_n(2\mu+1)$ are the shifted Legendre polynomials and shifted second kind Legendre functions, respectively. If we use Eq.(30) in Eq.(2), we obtain

$$Z_0(\mu,\omega) = 2\sum_{n=0}^{\infty}(2n+1)(-1)^n Q_n(2\mu+1)A_n(\omega), \qquad 0 \le (\omega,\mu) \le 1 \tag{31}$$

where $A_n(\omega)$ 's are defined as

$$A_n(\omega) = \int_0^1 P_n(2\mu'-1)H(\mu',\omega)d\mu', \qquad 0 \le \omega \le 1 \tag{32}$$

When Eq.(31) is used in Eq.(21), $A_n(\omega)$ functions that will satisfy the differential equation can be found easily. The formal series of $Q_n(2\mu+1)$ Legendre functions must be used in the process of obtaining the $A_n(\omega)$ functions. Formal series of Legendre functions $Q_n(2\mu+1)$ can be given as(Bell, 1968: page 91)

$$Q_n(2\mu+1) = \frac{2^n}{(2\mu+1)^{n+1}} \sum_{k=0}^{\infty} \frac{(n+k)!(n+2k)!}{k!(2n+2k+1)!(2\mu+1)^{2k}}, \qquad 0 \le \mu \le 1 \tag{33}$$

Let's use Eq.(33) and Eq.(31) in Eq.(21) and when we expand the result to the series with respect to $1/\mu$, we encountered an interesting situation; it turns out that all odd-numbered $A_n(\omega)$ functions are zero, that is

$$A_{2n+1}(\omega)=0, \qquad n\geq 0, \qquad 0\leq \omega \leq 1, \tag{34}$$

Several even-numbered $A_n(\omega)$ functions, including the $A_0(\omega)$ function, can be given as follows:

$$\frac{\omega}{4}A_0^2(\omega)-A_0(\omega)+1=0, \qquad 0\leq\omega\leq 1 \tag{35}$$

$$A_2(\omega)=\frac{\omega A_0^2(\omega)}{4(\omega A_0(\omega)-6)}=\frac{A_0(\omega)-1}{(\omega A_0(\omega)-6)}, \qquad 0\leq\omega\leq 1, \tag{36}$$

$$A_4(\omega)=\frac{9(1-A_0(\omega))+\omega A_0(\omega)+\dfrac{9\omega}{4}}{(\omega A_0(\omega)-6)^2(\omega A_0(\omega)-10)}, \qquad 0\leq\omega\leq 1, \tag{37}$$

As can be seen from the equations above, it is not possible to write a recurrence relation for $A_{2n}(\omega)$. Readers may consider this an open problem. At least, we have a differential equation with which we can easily calculate as many $A_{2n}(\omega)$'s as desired. Approximate results for $Z_0(\mu,\omega)$ can be found by using these $A_{2n}(\omega)$ 's in Eq.(31). In this study, we calculated up to $A_{10}(\omega)$ and found that the results are the same as those found in the exact solution. Our purpose in presenting this section is to demonstrate the accuracy of the exact solution we will provide in the next section. Previous work ( Anli, Öztürk 2021) has shown that the second type of Legendre function yields very accurate results.

**Exact Solution H Function:** Finding the exact solutions of the Riccati equations depends on finding the particular solutions of them. If a particular solution cannot be found, what needs to be done is that (Rivera-Oliva 2025);

$$Z_0(\mu,\omega)=\frac{4}{\omega G(\mu,\omega)}\frac{\partial G(\mu,\omega)}{\partial\mu}, \qquad 0\leq(\omega,\mu)\leq 1, \tag{38}$$

where we define a new function $G(\mu,\omega)$ that satisfy Eq.(21). If we use Eq.(38) in Eq.(21), we obtain a second-order differential equation with variable coefficient;

$$\frac{\partial^2 G(\mu,\omega)}{\partial\mu^2}+\frac{\omega}{4\mu(1+\mu)}G(\mu,\omega)=0, \qquad 0\leq(\omega,\mu)\leq 1, \tag{39}$$

Solution of this equation is given by

$$G(\mu,\omega)=C_1(\omega)\,\mu^{\left(-\frac{1}{2}+\frac{\sqrt{1-\omega}}{2}\right)}(1+\mu)\ {}_2F_1\left(\left[\frac{3}{2}-\frac{\sqrt{1-\omega}}{2},\frac{1}{2}-\frac{\sqrt{1-\omega}}{2}\right],[1-\sqrt{1-\omega}],-\frac{1}{\mu}\right)+$$
$$C_2(\omega)\,\mu^{\left(-\frac{1}{2}-\frac{\sqrt{1-\omega}}{2}\right)}(1+\mu)\ {}_2F_1\left(\left[\frac{3}{2}+\frac{\sqrt{1-\omega}}{2},\frac{1}{2}+\frac{\sqrt{1-\omega}}{2}\right],[1+\sqrt{1-\omega}],-\frac{1}{\mu}\right),\ 0\le(\omega,\mu)\le 1 \tag{40}$$

where ${}_2F_1(\cdot)$ is the hypergeometric function. Since the equation is second order, it is expected that two independent solutions will emerge. However, the first equation gives negative results for some values of the variables $\mu$ and $\omega$. However, the H function must be positive (Chandrasekhar 1960). Therefore, the acceptable solution we seek is the second term in Eq. (40);

$$G(\mu,\omega)=\mu^{\left(-\frac{1}{2}-\frac{\sqrt{1-\omega}}{2}\right)}(1+\mu)\ {}_2F_1\left(\left[\frac{3}{2}+\frac{\sqrt{1-\omega}}{2},\frac{1}{2}+\frac{\sqrt{1-\omega}}{2}\right],[1+\sqrt{1-\omega}],-\frac{1}{\mu}\right),0\le(\omega,\mu)\le 1 \tag{41}$$

Since we will use Eq.(41) in Eq.(38), we did not write the constant $C_2(\omega)$ (it will be vanish), so there is no need to determine this constant. As a result, the exact solution of the Chandrasekhar H function for isotropic case can be given as follows

$$H(\mu,\omega)=\frac{G(\mu,\omega)}{G(\mu,\omega)-2\mu\dfrac{\partial G(\mu,\omega)}{\partial\mu}},\qquad 0\le(\omega,\mu)\le 1 \tag{42}$$

where $G(\mu,\omega)$ is given in Eq.(41).

## 4. Conclusions

Firstly, to obtain the exact solution of the Chandrasekhar H function, we obtained its differential equation. As mentioned in the third section, we used two different methods to solve the differential equation we obtained. Our goal in doing so was to verify the accuracy of our calculations. We observed that the results obtained by both methods were identical. All operations and mathematical derivations are handled visually in a mathematical software.

After some rather difficult mathematical processes, we found the exact solution of the Chandrasekhar H function for isotropic case. As it is understood from the studies done so far, the Chandrasekhar H function contains sensitive numbers and since the sensitivity of these numbers cannot be noticed enough on the graph, so we have given our results in tabular form. In the table we have prepared, we have presented Chandrasekhar's original results side by side to compare the results we obtained. When the results in the table are carefully examined, it is observed that the difference between the results for small values of $\omega$ is quite small, while the differences between the results for values of $\omega$ close to 1 are large. Numerically speaking,

Table I

H-Function obtained numerically

| $\mu$ | $\omega$=0.1 | | $\omega$=0.2 | | $\omega$=0.3 | | $\omega$ =0.4 | | $\omega$ =0.5 | |
|---|---|---|---|---|---|---|---|---|---|---|
| | CHR | Eq.(42) | CHR | Eq.(42) | CHR | Eq.(42) | CHR | Eq.(42) | CHR | Eq.(42) |
| 0.00 | 1.00000 | 1.00000 | 1.00000 | 1.00000 | 1.00000 | 1.00000 | 1.00000 | 1.00000 | 1.00000 | 1.00000 |
| 0.05 | 1.00783 | 1.00783 | 1.01608 | 1.01615 | 1.02484 | 1.02505 | 1.03422 | 1.03465 | 1.04439 | 1.04512 |
| 0.10 | 1.01238 | 1.01241 | 1.02562 | 1.02578 | 1.03989 | 1.04027 | 1.05535 | 1.05615 | 1.07241 | 1.07378 |
| 0.15 | 1.01584 | 1.01589 | 1.03295 | 1.03314 | 1.05155 | 1.05204 | 1.07196 | 1.07298 | 1.09474 | 1.09652 |
| 0.20 | 1.01864 | 1.01869 | 1.03892 | 1.03914 | 1.06115 | 1.06171 | 1.08577 | 1.08693 | 1.11349 | 1.11556 |
| 0.25 | 1.02099 | 1.02105 | 1.04396 | 1.04419 | 1.06930 | 1.06991 | 1.09758 | 1.09885 | 1.12968 | 1.13197 |
| 0.30 | 1.02300 | 1.02306 | 1.04829 | 1.04855 | 1.07637 | 1.07701 | 1.10789 | 1.10924 | 1.14391 | 1.14637 |
| 0.35 | 1.02475 | 1.02482 | 1.05209 | 1.05235 | 1.08259 | 1.08325 | 1.11700 | 1.11842 | 1.15659 | 1.15917 |
| 0.40 | 1.02630 | 1.02636 | 1.05546 | 1.05572 | 1.08811 | 1.08880 | 1.12516 | 1.12661 | 1.16800 | 1.17067 |
| 0.45 | 1.02768 | 1.02774 | 1.05847 | 1.05874 | 1.09308 | 1.09377 | 1.13251 | 1.13399 | 1.17833 | 1.18107 |
| 0.50 | 1.02892 | 1.02898 | 1.06117 | 1.06145 | 1.09756 | 1.09827 | 1.13918 | 1.14068 | 1.18776 | 1.19054 |
| 0.55 | 1.03004 | 1.03010 | 1.06363 | 1.06390 | 1.10164 | 1.10235 | 1.14528 | 1.14678 | 1.19640 | 1.19922 |
| 0.60 | 1.03106 | 1.03112 | 1.06587 | 1.06615 | 1.10538 | 1.10608 | 1.15087 | 1.15238 | 1.20436 | 1.20721 |
| 0.65 | 1.03199 | 1.03205 | 1.06793 | 1.06820 | 1.10881 | 1.10951 | 1.15602 | 1.15753 | 1.21173 | 1.21459 |
| 0.70 | 1.03284 | 1.03290 | 1.06982 | 1.07009 | 1.11198 | 1.11268 | 1.16080 | 1.16230 | 1.21858 | 1.22143 |
| 0.75 | 1.03363 | 1.03369 | 1.07157 | 1.07183 | 1.11491 | 1.11560 | 1.16523 | 1.16672 | 1.22495 | 1.22780 |
| 0.80 | 1.03436 | 1.03442 | 1.07319 | 1.07345 | 1.11763 | 1.11832 | 1.16935 | 1.17084 | 1.23091 | 1.23375 |
| 0.85 | 1.03504 | 1.03510 | 1.07469 | 1.07495 | 1.12017 | 1.12085 | 1.17320 | 1.17468 | 1.23648 | 1.23931 |
| 0.90 | 1.03567 | 1.03573 | 1.07610 | 1.07635 | 1.12254 | 1.12322 | 1.17681 | 1.17827 | 1.24171 | 1.24453 |
| 0.95 | 1.03626 | 1.03632 | 1.07741 | 1.07767 | 1.12476 | 1.12543 | 1.18019 | 1.18165 | 1.24664 | 1.24943 |
| 1.00 | 1.03682 | 1.03687 | 1.07864 | 1.07889 | 1.12685 | 1.12751 | 1.18337 | 1.18482 | 1.25128 | 1.25405 |

CHR: Chandrasekhar's Results (in his book)

Table I(continued)

| $\mu$ | $\omega$ =0.6 | | $\omega$ =0.7 | | $\omega$ =0.8 | | $\omega$ =0.9 | | $\omega$ =1.0 | |
|---|---|---|---|---|---|---|---|---|---|---|
| | CHR | Eq.(42) | CHR | Eq.(42) | CHR | Eq.(42) | CHR | Eq.(42) | CHR | Eq.(42) |
| 0.00 | 1.00000 | 1.00000 | 1.00000 | 1.00000 | 1.00000 | 1.00000 | 1.00000 | 1.00000 | 1.00000 | 1.00000 |
| 0.05 | 1.05544 | 1.05673 | 1.06780 | 1.06990 | 1.08200 | 1.08548 | 1.09990 | 1.10558 | 1.13680 | 1.15199 |
| 0.10 | 1.09137 | 1.09371 | 1.11306 | 1.11688 | 1.13880 | 1.14506 | 1.17220 | 1.18285 | 1.24740 | 1.27767 |
| 0.15 | 1.12045 | 1.12351 | 1.15036 | 1.15541 | 1.18660 | 1.19504 | 1.23490 | 1.24970 | 1.35080 | 1.39576 |
| 0.20 | 1.14517 | 1.14877 | 1.18253 | 1.18854 | 1.22860 | 1.23880 | 1.29140 | 1.30973 | 1.45030 | 1.50969 |
| 0.25 | 1.16674 | 1.17075 | 1.21095 | 1.21772 | 1.26630 | 1.27794 | 1.34330 | 1.36463 | 1.54730 | 1.62092 |
| 0.30 | 1.18587 | 1.19020 | 1.23643 | 1.24382 | 1.30060 | 1.31342 | 1.39140 | 1.41539 | 1.64250 | 1.73027 |
| 0.35 | 1.20304 | 1.20762 | 1.25951 | 1.26741 | 1.33200 | 1.34588 | 1.43640 | 1.46269 | 1.73640 | 1.83821 |
| 0.40 | 1.21861 | 1.22337 | 1.28063 | 1.28891 | 1.36110 | 1.37579 | 1.47850 | 1.50699 | 1.82930 | 1.94507 |
| 0.45 | 1.23280 | 1.23771 | 1.30003 | 1.30863 | 1.38810 | 1.40350 | 1.51830 | 1.54868 | 1.92130 | 2.05106 |
| 0.50 | 1.24581 | 1.25084 | 1.31796 | 1.32681 | 1.41320 | 1.42929 | 1.55600 | 1.58803 | 2.01280 | 2.15637 |
| 0.55 | 1.25781 | 1.26292 | 1.33459 | 1.34366 | 1.43680 | 1.45338 | 1.59180 | 1.62528 | 2.10370 | 2.26109 |
| 0.60 | 1.26893 | 1.27409 | 1.35009 | 1.35931 | 1.45900 | 1.47596 | 1.62590 | 1.66063 | 2.19410 | 2.36534 |
| 0.65 | 1.27925 | 1.28446 | 1.36457 | 1.37392 | 1.47980 | 1.49717 | 1.65830 | 1.69424 | 2.28420 | 2.46918 |
| 0.70 | 1.28888 | 1.29411 | 1.37815 | 1.38759 | 1.49950 | 1.51716 | 1.68930 | 1.72627 | 2.37400 | 2.57267 |
| 0.75 | 1.29788 | 1.30312 | 1.39090 | 1.40041 | 1.51820 | 1.53603 | 1.71900 | 1.75683 | 2.46350 | 2.67586 |
| 0.80 | 1.30631 | 1.31156 | 1.40291 | 1.41247 | 1.53580 | 1.55389 | 1.74740 | 1.78603 | 2.55270 | 2.77879 |
| 0.85 | 1.31424 | 1.31948 | 1.41425 | 1.42383 | 1.55260 | 1.57081 | 1.77460 | 1.81398 | 2.64170 | 2.88148 |
| 0.90 | 1.32171 | 1.32694 | 1.42497 | 1.43456 | 1.56850 | 1.58688 | 1.80080 | 1.84076 | 2.73060 | 2.98398 |
| 0.95 | 1.32875 | 1.33396 | 1.43512 | 1.44471 | 1.58370 | 1.60216 | 1.82590 | 1.86644 | 2.81930 | 3.08629 |
| 1.00 | 1.33541 | 1.34059 | 1.44476 | 1.45433 | 1.59820 | 1.61671 | 1.85010 | 1.89110 | 2.90780 | 3.18844 |

CHR: Chandrasekhar's Results (in his book)

when $\omega$ is small, there is an average difference of 0.02% between the results, while when $\omega = 1.0$, the differences between the results are observed to be around 9% on average.

We obtained our results using the exact functional expression we obtained in the solution. Therefore, the difference between the results may be due to the method or approach of Chandrasekhar used for the numerical solution.

Another parameter related to this subject is the moments of the H function. In the studies conducted on this subject so far, only the exact functional expression for the moment $\alpha_0(\omega)$ has been given in literature. Functional expressions for the other moments could have not been given. In this study, functional expressions for all moments have been derived (see Eq.(26)-Eq. (29)). Thus, we will not give a table or graph regarding the numerical calculation of the moments of H function.

It is necessary to examine whether the solutions we obtained provide boundary values for the H function. Using Eq.(1), the limit values of the H function at $\mu \to 0$ and $\mu \to \infty$ can be easily obtained as follows, respectively;

$$H(0,\omega) = 1, \qquad H(\infty,\omega) = \frac{1}{1 - \dfrac{\omega}{2}\alpha_0(\omega)} = \frac{1}{\sqrt{1-\omega}}, \qquad 0 \le \omega \le 1 \tag{43}$$

The limiting value of the H function at $\mu \to \infty$ in Eq.(1) or Eq. (43) has no physical or mathematical meaning because $\mu$ cannot exceed 1. It is only functionally meaningful and can be used to understand the functional behavior H function. Eq.(31) and Eq.(38) satisfy exactly Eq.(43) in both limit cases of $\mu$.

**Declaration of competing interest:** The author declares that he has no conflict of interest.